\documentclass[conference]{IEEEtran}
\IEEEoverridecommandlockouts
\usepackage{url}
\usepackage{xurl}
\usepackage{cite}
\usepackage{amsmath,amssymb,amsfonts}
\usepackage{algorithmic}
\usepackage{graphicx}
\usepackage{textcomp}
\usepackage{xcolor}
\usepackage{hyperref}
\usepackage{stfloats}
\def\BibTeX{{\rm B\kern-.05em{\sc i\kern-.025em b}\kern-.08em
  T\kern-.1667em\lower.7ex\hbox{E}\kern-.125emX}}
\begin{document}

\title{Analysis-Driven Procedural Generation of an Engine Sound Dataset with Embedded Control Annotations%
\thanks{
Dataset: \url{https://doi.org/10.5281/zenodo.16883336} and \\ \url{https://huggingface.co/datasets/rdoerfler/procedural-engine-sounds}. \\
Analysis code: \url{https://github.com/rdoerfler/engine-order-analysis}
}
}

\author{\IEEEauthorblockN{Robin Doerfler}
\IEEEauthorblockA{\textit{Audio Research \& Development} \\
\textit{Impulse Audio Lab GmbH}\\
Munich, Germany \\
robin.doerfler@impulse-audio-lab.com}
\and
\IEEEauthorblockN{Lonce Wyse}
\IEEEauthorblockA{\textit{Music Technology Group} \\
\textit{Universitat Pompeu Fabra}\\
Barcelona, Spain \\
lonce.wyse@upf.edu}
}

\maketitle

\begin{abstract}
Computational engine sound modeling is central to the automotive audio industry, particularly for active sound design applications and virtual prototyping. Emerging data-driven engine sound synthesis methods require large volumes of standardized, clean audio recordings with precisely time-aligned operating-state annotations: data that is difficult to obtain due to high costs, specialized measurement equipment requirements, and inevitable noise contamination. We present an analysis-driven framework for generating engine audio with sample-accurate control annotations. The method extracts harmonic structures from real recordings through pitch-adaptive spectral analysis, which then drive an extended parametric harmonic-plus-noise synthesizer. With this framework, we augment 5--10\,min of source audio per engine 15--30$\times$ via diverse control trajectories and parametric variation, producing the \textit{Procedural Engine Sounds Dataset} (19.0~h, 5{,}935 files): a set of engine audio signals with sample-accurate RPM and torque annotations spanning a wide range of operating conditions, signal complexities, and harmonic profiles. Comparison against real recordings validates that the synthesized data preserves characteristic harmonic structures, and a baseline differentiable synthesis network trained on the dataset confirms its suitability for data-driven engine sound modeling. The dataset is released publicly to support research on engine timbre analysis, control parameter estimation, and neural generative synthesis.
\end{abstract}


\section{Introduction}
\label{sec:intro}
Accurate modeling of engine acoustics is central to traditional noise--vibration--harshness control and virtual prototyping \cite{baldanPhysicallyInformedCar2015, heitbrinkDesignDrivingSimulation2007}, as well as to modern active sound design applications \cite{afmalmborgEvaluationCarEngine2021, banSynthesisCarNoise2002, chenSynthesisingSoundCar2021, dupreAnalysisSynthesisEngine2023, lanslotsComprehensiveAutomotiveActive2020, moonActiveSoundDesign2020} and emerging data-driven synthesis approaches \cite{liRealTimeAutomotiveEngine2024, lundbergDataDrivenProceduralAudio2020, lobatoMotor2SynthLeveragingDifferentiable2025, doerflerPhysicsInformedNeuralEngine2026}.
Many of these applications, particularly machine-learning methods, require large sets of audio signals that are free from confounding noise and accompanied by precisely time-aligned annotations of engine operating parameters.

However, such datasets are scarce: vehicle recordings are expensive to obtain and inevitably contaminated by environmental and mechanical noise, while accessing ground-truth operating parameters often requires proprietary equipment.

Existing public datasets primarily target classification or detection tasks, and consist of real-world recordings with varying recording conditions and only coarse or missing temporal annotations \cite{gemmekeAudioSetOntology2017, akbalVehicleInteriorSound2022, dohiMIMIIDGSound2022, blahutCarSoundDataset2023, grollmischElectricalMotorSound2024}. Furthermore, recorded data cannot be systematically augmented or modified under controlled conditions, limiting the ability to evaluate algorithms across precisely defined scenarios. Procedural synthesis can generate such controllable, scalable, and perfectly annotated corpora \cite{engelNeuralAudioSynthesis2017}, but requires faithful reproduction of real acoustics for practical applicability.

This paper proposes a signal-processing framework for analysis-driven procedural synthesis of engine sounds. The dominant spectral characteristics of engine sounds are defined by their harmonic structure, described in terms of \textit{engine orders}: sinusoidal components at integer and half-integer multiples of the crankshaft rotation frequency, expressed in revolutions per minute (RPM). The half-integer spacing arises because, in four-stroke engines, each cylinder fires every two crankshaft revolutions.

The framework extracts the positional deviation and magnitude of each order as functions of RPM and torque from engine recordings, using a spectral analysis approach that combines pitch-adaptive resampling, frequency-aligned FFT analysis, and centroid-based order tracking. The extracted parameters drive a harmonic-plus-noise synthesizer with resonator modeling, enabling controlled generation of realistic engine sound signals with adjustable operating conditions. Control variables (RPM and torque) are embedded directly into the audio stream, providing sample-accurate ground-truth annotations without requiring external metadata files.

Using the proposed framework, we generate the Procedural Engine Sounds Dataset (19.0\,h, 5{,}935 files) covering diverse operating conditions under noise-free, fully synchronized settings. The main contributions are: (i) an analysis-driven framework for procedurally generating and augmenting engine audio with sample-accurate control annotations, including controllable parametric modification to produce meaningful graded signal complexity; (ii) the resulting publicly released dataset; and (iii) validation that the synthesized signals preserve characteristic engine-order structure across operating conditions, and the augmentation provides sufficient scale and operating-state coverage for data-driven engine sound modeling.

\section{Signal Processing Framework}
\label{sec:framework}

The proposed framework integrates three main components: (1) spectral analysis for extracting performance-parameter-dependent harmonic features from real recordings, (2) parametric synthesis for controlled signal generation, and (3) synchronized multi-channel encoding for sample-accurate ground truth embedding. This enables expansion of limited real-world recordings into clean, large-scale annotated datasets with sample-accurate ground truth labels, while maintaining acoustic authenticity.

\subsection{Spectral Analysis Pipeline for Feature Extraction}
\label{sec:analysis}
The synthesis parameters derive from systematic spectral analysis of real-world engine recordings from various vehicle configurations.

\subsubsection{Angle-Domain Resampling}
The audio content, sampled at 16 kHz, was segmented into frames of 65{,}536 samples (4.096-second chunks); frames exhibiting zero RPM (engine off) were excluded from analysis.

Let the instantaneous fundamental frequency $f_0 = \text{RPM}/60$ denote the crankshaft rotation frequency in Hz. To prevent orders from drifting between FFT bins as RPM varies, each frame was time-warped into a domain where $f_0$ is held constant at the frame-mean value $\overline{f_0}$, effectively resampling the signal at uniform crankshaft-angle increments rather than uniform time~\cite{ saavedraAccurateAssessmentComputed2006}.
Given the discrete envelope $f_0[n]$, faster-than-average intervals are stretched and slower ones compressed, so the read position in the original signal for each output sample is the cumulative sum of the per-step ratios $\overline{f_0}/f_0[i]$:
\begin{equation}
t'[n] = \sum_{i=0}^{n-1} \frac{\overline{f_0}}{f_0[i]}, \qquad t'[0] = 0,
\end{equation}
where $n$ indexes the output samples and $t'[n]$ is the (continuous-valued) position in the original signal from which to read. The original signal is evaluated at these fractional positions via cubic-spline interpolation, which provides smoother reconstruction than linear resampling:
\begin{equation}
y_{\text{repitched}}[n] = \text{CubicSpline}(y_{\text{original}},\, t'[n]).
\end{equation}

\subsubsection{Frequency-Aligned FFT Analysis}
Per-order features extracted at different RPMs are only directly comparable if the analysis presents each order under identical conditions. A fixed FFT size cannot deliver this: as $f_0$ varies, each order lands at a different fractional bin position, surrounded by different spectral leakage patterns. We instead scale the FFT size with $f_0$ so that the bin grid stretches and contracts in proportion to the engine's fundamental, locking each order to a fixed bin index.

Given the sampling rate $f_s$, the Blackman analysis window of length $M$ spans $P$ fundamental periods, and the FFT size $N_{\text{FFT}}$ is obtained by zero-padding with factor $z$:
\begin{equation}
M = \left\lfloor \frac{f_s}{f_0} \cdot P \right\rceil, \qquad N_{\text{FFT}} = M \cdot z,
\end{equation}
yielding a bin spacing of
\begin{equation}
\Delta f = \frac{f_0}{P \cdot z}.
\end{equation}
We use $P = 20$, keeping the Blackman mainlobe narrow relative to the inter-order spacing to suppress leakage between adjacent orders, and $z = 4$ to densely sample each mainlobe, stabilizing the centroid and parabolic peak estimates described in the next subsection.
This places $P \cdot z = 80$ bins per integer engine order at every RPM: the crankshaft frequency $f_0$ sits at bin $b_{1.0} = 80$, and order $h \in \{0.5, 1.0, 1.5, 2.0 \ldots, 64.0\}$ at bin $b_h = h \cdot b_{1.0}$. Each order therefore occupies the center of an identical 40-bin region with the same mainlobe shape and the same distance to its neighbors, so differences between orders or RPMs reflect signal content rather than grid geometry.

\subsubsection{Centroid-Based Order Estimation}
Real engine orders generally deviate from their nominal positions due to mechanical coupling, combustion irregularities, thermodynamic effects and structural resonances. The bin positions $b_h$ therefore represent ideal locations; the true positions $\hat{b}_h$ are estimated by taking the spectral centroid within a local region around each $b_h$.

For each order $h$, the analysis region $R_h$ extends halfway to the adjacent half-integer orders:
\begin{equation}
R_h = \left\{ k \in \mathbb{Z} \;:\; \left\lfloor \tfrac{b_{h-0.5} + b_h}{2} \right\rfloor \le k \le \left\lfloor \tfrac{b_h + b_{h+0.5}}{2} \right\rfloor \right\}.
\end{equation}
A window $w_h[k]$ with unity gain in the center and smooth tapered edges is applied to the magnitude spectrum $|X[k]|$, suppressing leakage from neighboring orders while preserving the symmetry around $b_h$ that an unbiased centroid requires. The weighted centroid is
\begin{equation}
\hat{b}_h = \frac{\displaystyle\sum_{k \in R_h} k \cdot |X[k]| \cdot w_h[k]}{\displaystyle\sum_{k \in R_h} |X[k]| \cdot w_h[k]},
\end{equation}
giving a fractional bin position $\hat{b}_h$, whose magnitude \text{$A_h = |X[\hat{b}_h]|$} is recovered via three-point parabolic interpolation~\cite{smithSpectralAudioSignal2011}. Centroid estimation is less susceptible to spurious peaks than direct peak picking, particularly when neighboring orders or noise place a local maximum off the true order frequency.

Per-order deviations are quantified as the order-normalized offset between estimated and ideal positions:
\begin{equation}
\label{eq:h_deviation}
\delta_h = \frac{\hat{b}_h}{b_{1.0}} - h,
\end{equation}
where positive $\delta_h$ indicates a shift toward higher frequencies (sharp relative to the nominal order) and negative the reverse. Computed across all frames, the pair $(\delta_h, A_h)$ forms the per-order acoustic fingerprint of each engine configuration as a function of RPM and torque.

\subsection{Parametric Synthesis Model}
\label{sec:synthesis}

\subsubsection{Real-Time Synthesis Architecture}
The synthesizer reconstructs engine sound from the per-order fingerprint $(\delta_h, A_h)$ extracted in Section~\ref{sec:analysis}, augmented with stochastic and resonant components. One sine oscillator is instantiated per order $h \in \{0.5, 1.0, \ldots, 64.0\}$, giving 128 oscillators in total. At each synthesis step, the current operating point $(\text{RPM}(t), \text{Torque}(t))$ indexes the stored $(\delta_h, A_h)$ tables via bilinear interpolation, yielding time-varying $A_h(t)$ and $\delta_h(t)$ that drive the oscillator bank.

\subsubsection{Harmonic Synthesis}
The oscillator bank generates time-domain audio through additive synthesis:
\begin{equation}
x(t) = \sum_{h} A_h(t) \sin\!\left(2\pi \int_0^t f_h(\tau)\, d\tau\right),
\end{equation}
where the phase integral accumulates the instantaneous frequency of each order over time. The frequency $f_h(t)$ is the crankshaft frequency scaled by the nominal order number and detuned by the stored deviation:
\begin{equation}
f_h(t) = \bigl(h+\delta_h(t)\bigr) \cdot f_0(t).
\end{equation}
\subsubsection{Noise and Resonator Components}
The extracted fingerprint captures the deterministic harmonic structure but not the stochastic and resonant content that gives a real engine its texture. Two augmentation components recover these:

\paragraph{Noise Synthesis}

Pink noise amplitude-modulates the harmonic sum, perturbing per-order amplitudes to emulate cycle-to-cycle combustion variability:
\begin{equation}
n_{\text{turbulence}}(t) = x(t) \cdot [1 - \alpha + \alpha \cdot n_{\text{pink}}(t)], \quad 0 \leq \alpha \leq 1,
\end{equation}
where $n_{\text{pink}}(t)$ is normalized pink noise and $\alpha$ controls the modulation depth.

Impulsive noise from valve events and intake resonances is synthesized as filtered white noise amplitude-modulated by envelopes locked to low-order crankshaft oscillations:
\begin{equation}
n_{\text{burst}}(t) = \text{LPF}(n_\text{white}(t)) \cdot \sum_{m=1}^{4} w_m |\phi_m(t)|^{\gamma_m},
\end{equation}
where $n_{\text{white}}(t)$ is white noise, $\phi_m(t) = \sin\!\left(2\pi h_m \int_0^t f_0(\tau)\,d\tau\right)$ with $h_m \in \{0.5, 1.0, 1.5, 2.0\}$, $w_m$ and $\gamma_m$ are per-component mixing weights and shaping exponents, and LPF denotes an 18 dB/oct low-pass filter. The $w_m$, $\gamma_m$, and filter cutoff are tuned heuristically per engine configuration.

The noise sources and resonator banks are instantiated independently for the left and right channels, producing two decorrelated realizations of the harmonic model --- analogous to two exhaust paths sharing the same cylinder pressures but radiating independently.

\paragraph{Resonator Modeling}
Exhaust resonances are modeled as a parallel bank of Karplus--Strong-style resonators~\cite{smithPhysicalAudioSignal}:
\begin{equation}
y_k(t) = s(t) + g_k \cdot y_k(t - \tau_k), \qquad y_r(t) = \sum_{k=1}^{N_r} y_k(t),
\end{equation}
where $s(t) = x(t) + n_{\text{turbulence}}(t) + n_{\text{burst}}(t)$ is the shared input and each branch has its own feedback gain $g_k < 1$ and delay $\tau_k$ tuned to a characteristic exhaust resonance. A shared damping filter inside each feedback loop further controls the high-frequency decay of the resonator bank uniformly.

The per-branch $g_k$ and $\tau_k$, the damping characteristics, and the resonator count $N_r$ are exposed as user-facing parameters for timbral shaping. Initial values for $\tau_k$ are set from typical exhaust pipe and chamber lengths, with $g_k$ and damping then empirically adjusted to match reference recordings.

\subsection{Synchronized Multi-Channel Encoding}
\label{sec:encoding}

The system generates synchronized four-channel audio sampled at 48 kHz: stereo engine audio (channels 1--2) with time-aligned engine seed (rotational frequency in RPM) and torque (rotational force in newton-meters) control parameters (channels 3--4). RPM and torque were measured directly during recording via instrumented vehicle telemetry. Control parameters are normalized to [-1, 1] using fixed boundaries of 10,000 RPM and 1,000 Nm, then encoded at 16-bit resolution in dedicated audio channels. This provides 0.3 RPM and 0.03 Nm resolution, enabling sample-accurate reconstruction of operating conditions directly from the audio stream without external metadata files.

\section{Dataset Generation and Operational Coverage} \label{sec:dataset}

The dataset is derived from recordings of performance-class production vehicles spanning V8, inline-6, and inline-4 petrol engine configurations with varied exhaust systems. Harmonic timbre features were extracted from 5--10 minutes of audio per vehicle. Performance parameter control traces (RPM and torque sequences) were pooled from 2.5 hours of recordings across various vehicles and engine configurations under normal and dynamic driving scenarios, as well as systematic engine operation-state sampling on a dynamometer. Statistical analysis confirms comprehensive operational coverage: RPM ranges from 0 to 7{,}007 (mean: 3{,}171, std: 1{,}714), and torque from -107 to 718 Nm (mean: 120, std: 201), encompassing acceleration, cruising, deceleration, gear shifts, and idle operation.

By expanding timbre features from each source vehicle across the full control-trace pool, we achieve 15--30$\times$ data augmentation. Together with parametric variation in resonator and noise characteristics, this yields 5{,}935 files ($\sim$19.0 hours, 24.5 GB) of procedurally generated engine audio organized into eight acoustically distinct subsets, with clips up to 12.3\,s (minimum 4.1\,s for residual chunks, to support sequence-modeling tasks). In combination with the breadth of control sequences, this dataset provides standardized yet diverse data for systematic development and evaluation of synthesis and parameter estimation algorithms. The embedded control traces in the four-channel format (Section~\ref{sec:encoding}) further enable researchers to generate extended datasets using the proposed framework or comparable synthesis approaches with their own recordings.

\section{Validation}
\label{sec:validation}

We assess the proposed framework along three dimensions: (1) acoustic authenticity when augmenting limited recordings to large-scale corpora, (2) usability of the resulting data for data-driven research, and (3) controllable, graded complexity across parametric modifications.

\begin{figure*}[t]
\centering
\includegraphics[width=0.95\textwidth]{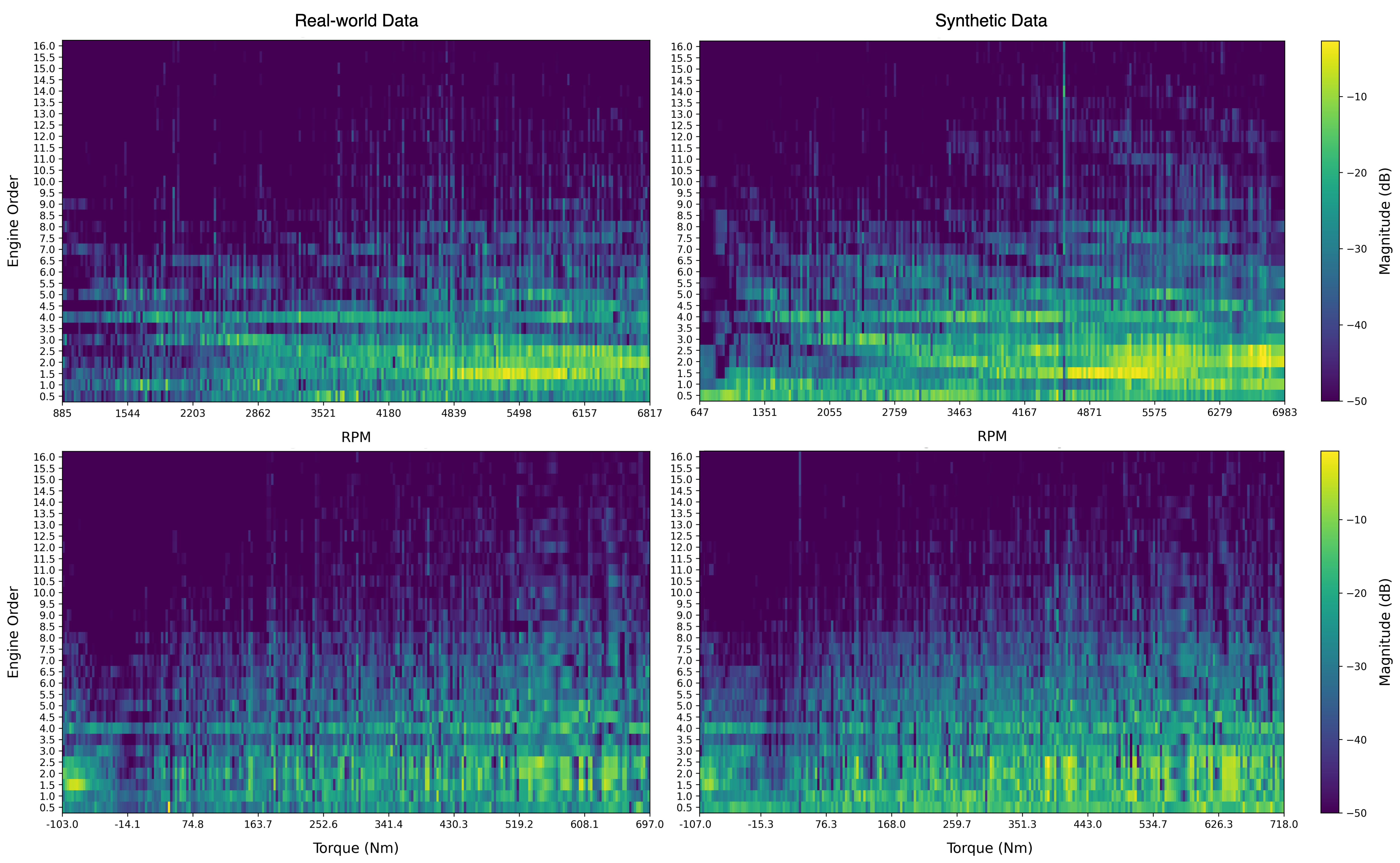}
\caption{Engine order magnitudes across RPM and torque range for one representative configuration, comparing source recordings (left, 90~min) to framework-generated synthetic signals (right, 150~min from 5~min extraction material). Upper panels plot per-order magnitude $A_h$ as a function of RPM; lower panels plot the same against torque. Preserved structural features demonstrate acoustic authenticity under 30$\times$ augmentation.}
\label{fig:comparison}
\end{figure*}

\subsection{Framework Effectiveness: Analysis-Driven Augmentation}
We compare order magnitude profiles $A_h$ between source recordings and resynthesized signals to assess whether limited-source analysis yields representative harmonic structures across unseen control traces. Figure~\ref{fig:comparison} shows one representative configuration; agreement holds across all eight subsets. Because the framework discards source noise and substitutes parameterized noise and resonator components for controllable timbral variation, exact spectral recreation is neither targeted nor a meaningful success criterion; we instead verify that fundamental acoustic behavior tracks operating state. Engine-specific signatures are preserved (dominant 4th order at V8 firing frequency, 1.5th order during engine-braking), with magnitude evolution corresponding across the RPM-torque operating space. Variations in higher orders ($>$8) reflect parametric modifications that extend timbral diversity beyond source material.

\subsection{Application Validation: Enabling Data-Driven Research}
To demonstrate that framework-generated data supports data-driven modeling, we trained a differentiable harmonic-plus-noise synthesis network (1.4M parameters) that reconstructs audio solely from RPM and torque inputs; full architectural details are given in~\cite{doerflerNeuralEngineSound2025}. We evaluate training dynamics on three dataset subsets A (inline-4), B (V8 SUV), and C (V8 sports car), selected to span increasing signal complexity: from predominantly harmonic content (A) through moderate noise and resonator modification (B) to strong harmonic deviations and resonator transformation (C). The model maps control parameters through GRU layers to time-varying parameters for 128 harmonic voices and 256 noise bands, trained with multi-resolution STFT loss over 100 epochs.

\begin{figure}[t]
\centering
\includegraphics[width=0.48\textwidth]{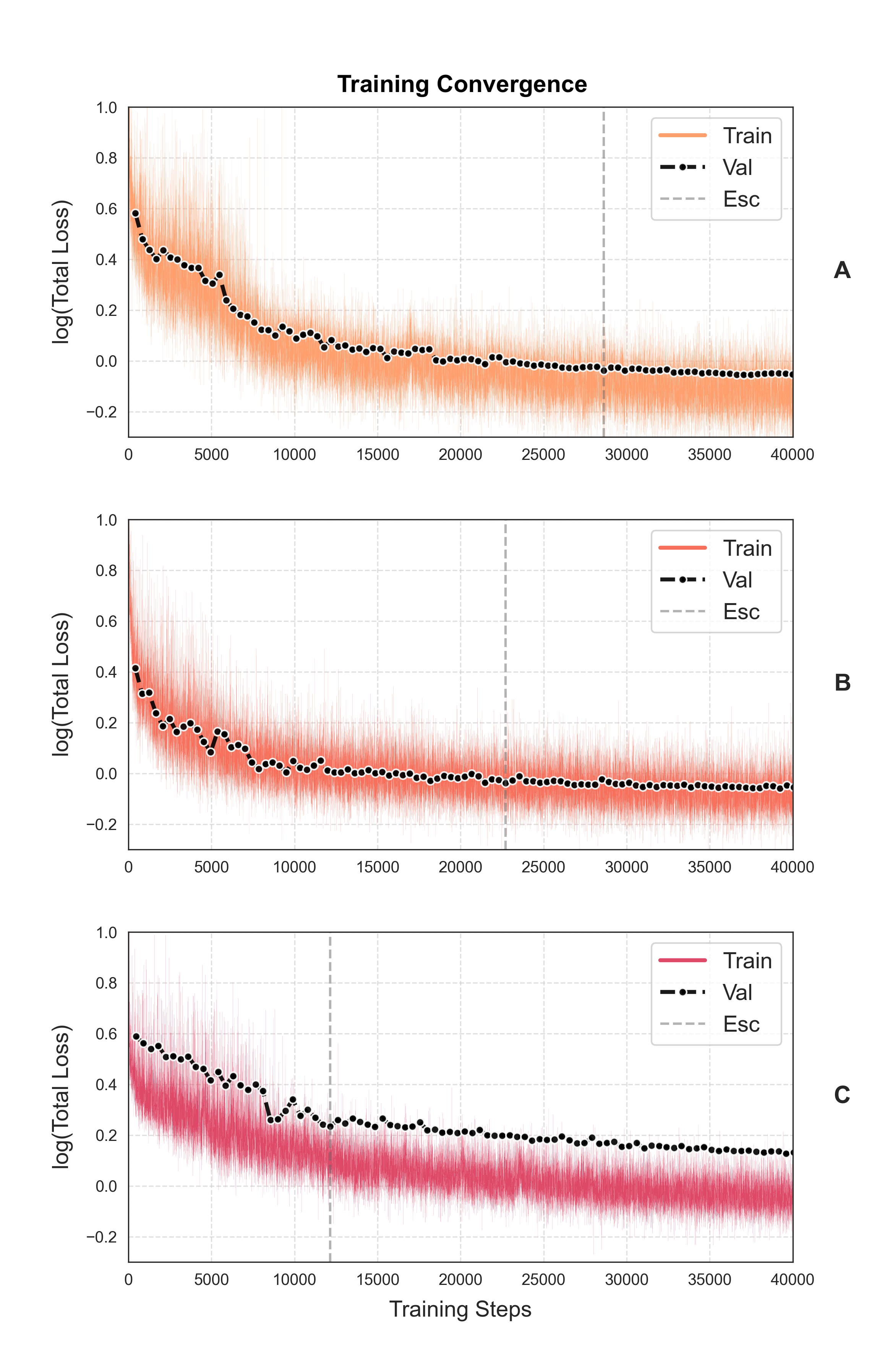}
\caption{Training and validation loss for the neural synthesis model on datasets A, B, and C. Stable convergence at early stopping demonstrates suitability for data-driven methods; Esc points (defined as positions with 8 following epochs of no significant improvement) shift earlier from A to C with increasing non-deterministic content.}
\label{fig:training}
\end{figure}

Figure~\ref{fig:training} shows stable convergence across all subsets with minimal train--validation gap at early stopping, evidencing that the augmented data provides the scale and operating-state coverage required for such training --- coverage that 5--10~min of source material alone does not afford. The RPM-torque annotations prove sufficient to drive reconstruction, supporting their use as physics-grounded control parameters. Esc markers shift progressively earlier from A to C, consistent with increased non-deterministic content (noise injection, resonator transformation) introducing irreducible variation in the RPM-torque-to-audio mapping --- meaningful timbral diversity rather than mere signal perturbation. Audio examples and comparative benchmark on these subsets available in~\cite{doerflerPhysicsInformedNeuralEngine2026}.

\section{Conclusion}
\label{sec:conclusion}

We presented an analysis-driven framework for procedural engine sound synthesis that generates annotated datasets from minimal source material, achieving 15--30$\times$ expansion while preserving engine-specific acoustic signatures. The released dataset, with sample-accurate embedded annotations, addresses documented limitations in existing engine audio resources and enables research on inverse parameter estimation (predicting RPM/torque from audio for automatic annotation and NVH diagnostics), data-driven synthesis development (learning parameter mappings without manual tuning), and systematic algorithm evaluation via controllable modifications. The documented analysis-synthesis pipeline lets researchers apply the framework to their own recordings for task-specific corpus generation, with code and dataset publicly available.

\bibliographystyle{IEEEtran}
\bibliography{refs}

\end{document}